\documentclass{ws-ijmpa}
\usepackage{url}
\usepackage[super,compress]{cite}
\usepackage{graphicx}
\usepackage{hyperref}
\hypersetup{colorlinks,urlcolor=black,citecolor=black,linkcolor=black,filecolor=black}
\usepackage{breakurl}

\begin{document}
\markboth{Feng-Zhi Chen and Xin-Qiang Li}{$CP$ asymmetries in $\tau\to K_S\pi\nu_\tau$ decays}

%
\catchline{}{}{}{}{}
%

\title{$CP$ asymmetries in $\tau\to K_S\pi\nu_\tau$ decays}

\author{Feng-Zhi Chen}

\address{Department of Physics, College of Physics $\&$ Optoelectronic Engineering,\\ Jinan University, Guangzhou 510632, P.R. China\\
fzchen@jnu.edu.cn}

\author{Xin-Qiang Li}

\address{Institute of Particle Physics and Key Laboratory of Quark and Lepton Physics~(MOE),\\ Central China Normal University, Wuhan, Hubei 430079, China\\
Center for High Energy Physics, Peking University, Beijing 100871, China\\	
xqli@mail.ccnu.edu.cn}

\maketitle

\begin{history}
\received{Day Month Year}
\revised{Day Month Year}
\end{history}

\begin{abstract}
We present here the CP asymmetries in the decay rate and angular distributions of $\tau\to K_S\pi\nu_\tau$ decays in the Standard Model (SM) and beyond (BSM). The CP asymmetries in the SM are induced by the CP violation in $K^0-\bar{K}^0$ mixing. To investigate the BSM CP-violating (CPV) effects, a model-independent analysis is performed by using the low-energy effective field theory (LEFT) framework at $\mu=2$~GeV. If one further assumes the BSM physics to stem from above the electroweak scale, the LEFT shall then be matched onto the SM effective field theory (SMEFT), the operators of which contributing to $\tau\to K_S\pi\nu_\tau$ decays will also contribute to the neutron electric dipole moment (EDM) and $D^0-\bar{D}^0$ mixing. The stringent bounds from the latter suggest that no remarkable CPV effects can be observed in either the decay rate or the angular distributions. The prospects for future measurements of these observables are also mentioned.

\end{abstract}

\ccode{PACS numbers: 14.60.Fg, 13.35.Dx, 12.15.Ji}


\section{Introduction}	

The hadronic decays of the $\tau$ lepton, besides serving as a clean laboratory for testing various low-energy aspects of the strong interaction~\cite{Pich:2013lsa,Pich:2024qob}, may also allow us to explore the non-standard CPV interactions~\cite{Bigi:2012kz,Kiers:2012fy}. 

In this presentation, we will focus on the CP asymmetries in $\tau\to K_S\pi\nu_\tau$ decays. Taking into account the CPV effect in $K^0-\bar{K}^0$ mixing, a non-zero indirect CP asymmetry in the SM can still arise in the processes involving a $K_S$ or a $K_L$ meson in the final state. Such an effect was first predicted in the decay rate~\cite{Bigi:2005ts}, and then in the angular distributions~\cite{Chen:2020uxi,Chen:2021udz} of $\tau\to K_S\pi\nu_\tau$ decays. Experimentally, the CP asymmetries in $\tau\to K_S\pi\nu_\tau$ decays have been searched for by CLEO~\cite{Anderson:1998ke,Bonvicini:2001xz}, Belle~\cite{Bischofberger:2011pw}, and BaBar~\cite{BaBar:2011pij}. The CLEO and Belle focused on the CP asymmetries in the angular distributions of the decays and obtained null results, while the BaBar concentrated on the CP asymmetry in the decay rate of the same processes and reported a non-zero CP asymmetry, which is however in conflict with the SM prediction~\cite{Bigi:2005ts,Grossman:2011zk} at a level of $2.8\sigma$. The latter has motivated many BSM explanations by including the extra contribution from non-standard tensor interactions~\cite{Cirigliano:2017tqn,Rendon:2019awg,Chen:2019vbr,Devi:2013gya,Dhargyal:2016kwp,Dhargyal:2016jgo,VonDetten:2021rax}, while only a few of studies~\cite{Bischofberger:2011pw,Chen:2021udz} focus on the former. It is found that only the vector-tensor interference contributes to the decay-rate asymmetry~\cite{Devi:2013gya}, while either the vector-scalar or the tensor-scalar interference has contribution to the angular distribution asymmetries~\cite{Chen:2021udz}. In each case, the $K\pi$ form factors play a key role in determining the amount of direct CP asymmetries in the decays. The $K\pi$ vector and scalar form factors have been well studied~\cite{Boito:2008fq,Boito:2010me,Jamin:2000wn,Jamin:2001zq,Jamin:2006tj}, and the tensor form factor is calculated by using the chiral theory with resonances (R$\chi$T)~\cite{Ecker:1988te,Ecker:1989yg} together with the dispersion relation~\cite{Cirigliano:2017tqn,Rendon:2019awg,Chen:2019vbr,VonDetten:2021rax}. The amount of CP asymmetry in the decay rate is suppressed~\cite{Cirigliano:2017tqn,Rendon:2019awg,Chen:2019vbr} due to the Watson's final-state interaction theorem~\cite{Watson:1954uc}. Nevertheless, since the vector-scalar and tensor-scalar phase differences are large enough, a search for direct CP asymmetries in the angular distributions of the decays is more promising~\cite{Chen:2020uxi,Chen:2021udz}.

The BSM CPV effects in $\tau\to K_S\pi\nu_\tau$ decays can be described in a model-independent way by using the $SU(3)_C\otimes U(1)_{em}$ invariant LEFT Lagrangian at a typical scale $\mu=2$~GeV. If the non-standard interactions originate from a weakly-coupled heavy new physics (NP) well above the electroweak scale, the $SU(2)_L$ invariance of the resulting SMEFT Lagrangian~\cite{Buchmuller:1985jz,Grzadkowski:2010es,Brivio:2017vri} would indicate that very strong limits on the imaginary parts of the scalar and tensor coefficients, $\mathrm{Im}[\hat{\epsilon}_S]$ and $\mathrm{Im}[\hat{\epsilon}_T]$, can also be obtained from the neutron EDM and the $D^0-\bar{D}^0$ mixing~\cite{Cirigliano:2017tqn,Chen:2021udz}. The stringent limit on $\mathrm{Im}[\hat{\epsilon}_T]$ has ruled out the possibility to reconcile the $2.8\sigma$ discrepancy in the decay-rate asymmetry. As for the BSM CP asymmetries in the angular distributions, unless there exist extraordinary cancellations between the NP contributions, neither the scalar nor the tensor interaction can produce any significant effects on the CP asymmetries (relative to the SM prediction) in the decays considered, especially under the ``single coefficient dominance'' assumption~\cite{Chen:2021udz}. 

\section{CP asymmetries in the decay-rate and angular distributions of \boldmath{$\tau\to K_S\pi\nu_\tau$ decays}}\label{sec:CPV}

Following the steps detailed in Refs.~\cite{Chen:2020uxi,Chen:2021udz}, the CP asymmetries in the decay rate and angular distributions of $\tau\to K_S\pi\nu_\tau$ decays are given, respectively, by
\begin{align}
A_\mathrm{rate}^{CP}\simeq& A^{CP}_\tau-A^{CP}_{K}\,,\label{eq:rateCPV}\\[0.2cm]
A_i^{CP}\simeq& \left(\langle\cos\alpha\rangle_i^{\tau^-}+\langle\cos\alpha\rangle_i^{\tau^+}\right)A^{CP}_K+\left(\langle\cos\alpha\rangle_i^{\tau^-}-\langle\cos\alpha\rangle_i^{\tau^+}\right)\,.\label{eq:angularCPV}
\end{align}
Here $A^{CP}_{K}\approx -2\mathrm{Re}(\epsilon_K)=-(3.32\pm0.06)\times10^{-3}$ represents the CP asymmetry in $K^0-\bar{K}^0$ mixing, with $\epsilon_K$ being the CPV parameter in the neutral kaon decays~\cite{Zyla:2020zbs}. The direct CP asymmetry in $\tau$ decays is defined as
\begin{align}
A^{CP}_\tau&=
\frac{\Gamma(\tau^+\to K^0\pi^+\bar \nu_\tau)-\Gamma(\tau^-\to \bar K^0\pi^-\nu_\tau)}
{\Gamma(\tau^+\to K^0\pi^+\bar \nu_\tau)+\Gamma(\tau^-\to \bar K^0\pi^-\nu_\tau)}\,.\label{eq:ACPTau}
\end{align}
The angular observables $\langle\cos\alpha\rangle_i^{\tau^{\pm}}$ denote the differential $\tau^{\pm}$ decay widths weighted by $\cos\alpha$ and are evaluated in the $i$-th bin of the $K\pi$ invariant mass squared, with
\begin{align}
\langle\cos\alpha\rangle_i^{\tau^-}\pm\langle\cos\alpha\rangle_i^{\tau^+}&=\frac{\int_{s_{1,i}}^{s_{2,i}}\int_{-1}^{1} \cos \alpha \left[\frac{d \Gamma^{\tau^-}}{d\omega}\pm\frac{d \Gamma^{\tau^+}}{d\omega}\right]d\omega}{\frac{1}{2}\int_{s_{1,i}}^{s_{2,i}}\int_{-1}^{1} \left[\frac{d \Gamma^{\tau^-}}{d\omega}+\frac{d \Gamma^{\tau^+}}{d\omega}\right]d\omega}\,,\label{eq:cossum}
\end{align}
where $\frac{d\Gamma^{\tau^\pm}}{d\omega}=\frac{d^2\Gamma(\tau^\pm\to K^0(\bar K^0)\pi^\pm\bar\nu_\tau(\nu_\tau))}{ds\,d\cos\alpha}$ are the doubly differential decay rates. 

Within the SM, we have $\Gamma^{\tau^+}=\Gamma^{\tau^-}$ and $\langle\cos\alpha\rangle_i^{\tau^-}=\langle\cos\alpha\rangle_i^{\tau^+}$, which then implies that $A_\mathrm{rate}^{CP}$ and $A_i^{CP}$ reduce separately to $A_{\mathrm{SM,rate}}^{CP}=-A^{CP}_K$ and $A_{\mathrm{SM},i}^{CP}=2\,\langle\cos\alpha\rangle_i^{\tau^-}A^{CP}_K$.
Numerically, $A_{\mathrm{SM,rate}}^{CP}=(3.32\pm0.06)\times10^{-3}$, and the predicted $A_{\mathrm{SM},i}^{CP}$ in the four different $K\pi$ invariant-mass bins are collected in Table~\ref{tab:belle}, which can be compared to the corresponding measurements provided by Belle~\cite{Bischofberger:2011pw}. In the presence of NP, however, it is generally expected that $\Gamma^{\tau^+}\neq\Gamma^{\tau^-}$ and $\langle\cos\alpha\rangle_i^{\tau^-}\neq\langle\cos\alpha\rangle_i^{\tau^+}$. As a consequence, one may obtain different values of $A_\mathrm{rate}^{CP}$ and $A_i^{CP}$ with respect to the SM expectations. A model-independent description of the decays in the LEFT framework will be discussed in next section.

\begin{table}[t]
	\tabcolsep 0.26in
	\renewcommand\arraystretch{1.2}
	\tbl{\small The SM predictions (the second column) and the Belle measurements (the third column) of the $CP$ asymmetries $A_{i}^{CP}$, and the observed number of signal events $n_i$ per mass bin divided by the number of total events $N_s$ (the fourth column) for four different $K\pi$ invariant-mass bins (the first column)~\cite{Bischofberger:2011pw}. \label{tab:belle} }
		{\begin{tabular}{cccc}
			\hline\hline
			$\sqrt{s}$~[GeV] & $A_{\mathrm{SM},i}^{CP}~[10^{-3}]$ & $A_{\mathrm{exp},i}^{CP}~[10^{-3}]$ & $n_i/N_s~[\%]$\\
			\hline
			$0.625-0.890$ & $0.39\pm0.01$ & $\phantom{-}7.9\pm3.0\pm2.8$ & $36.53\pm0.14$\\
			$0.890-1.110$ & $0.04\pm0.01$ &$\phantom{-}1.8\pm2.1\pm1.4$ & $57.85\pm0.15$\\
			$1.110-1.420$ & $0.12\pm0.02$ & $-4.6\pm7.2\pm1.7$ & $\phantom{0}4.87\pm0.04$\\
			$1.420-1.775$ & $0.27\pm0.05$ & $-2.3\pm19.1\pm5.5$ & $\phantom{0}0.75\pm0.02$\\
			\hline\hline
		\end{tabular}}
\end{table}

\section{LEFT analysis of \boldmath{$\tau^\pm\to K^0(\bar K^0)\pi^\pm\bar\nu_\tau(\nu_\tau)$} decays}
\label{sec:Effective}

Assuming the absence of other light degrees of freedom except for the SM ones below the electroweak scale, as well as the Lorentz and $SU(3)_C\otimes U(1)_{\text{em}}$ invariance, we can write the most general low-energy effective Lagrangian governing the $\tau^\pm\to K^0(\bar K^0)\pi^\pm\bar\nu_\tau(\nu_\tau)$ decays as~\cite{Cirigliano:2009wk,Bhattacharya:2011qm,Gonzalez-Solis:2019owk,Rendon:2019awg}
\begin{align}\label{eq:LEFT}
\mathcal{L}_\mathrm{eff}
=&-\frac{G^0_{F}V_{us}}{\sqrt{2}}\,(1+\epsilon_{L}+\epsilon_{R})\,\Big\{\bar{\tau}\gamma_{\mu}
(1-\gamma_{5})\nu_{\tau}\cdot\bar{u}\left[\gamma^{\mu}-(1-2\,\hat{\epsilon}_{R})\gamma^{\mu}\gamma_{5}\right]s\,\nonumber\\[0.1cm]
&+\bar{\tau}(1-\gamma_{5})\nu_{\tau}\cdot\bar{u}\left[\hat{\epsilon}_{S}-\hat{\epsilon}_{P}\gamma_{5}\right]s+
2\,\hat{\epsilon}_{T}\,\bar{\tau}\sigma_{\mu\nu}(1-\gamma_{5})\nu_{\tau}\cdot\bar{u}\sigma^{\mu\nu}s\Big\}+
\mathrm{h.c.}\,,
\end{align}
where the effective couplings $\hat{\epsilon}_{i}$ parametrize the non-standard NP contributions and can be generally complex, with the SM case recovered by setting all $\epsilon_{i}=0$. 

In terms of the NP parameters and the $K\pi$ form factors, the CP asymmetry in the decay rate and angular distributions can then be written, respectively, as~\cite{Cirigliano:2017tqn,Rendon:2019awg,Chen:2019vbr}
\begin{align}
A^{CP}_\tau
=&\frac{\mathrm{Im}[\hat{\epsilon}_{T}]\,G_{F}^{2}|V_{u s}|^{2} S_\mathrm{EW}}{128\,\pi^{3}\, m_{\tau}^{2}\, \Gamma(\tau \to K_S\pi\nu_{\tau})}\,\int_{s_{K\pi }}^{m_{\tau}^{2}} d s\left(1-\frac{m_{\tau}^{2}}{s}\right)^{2}\,\lambda^{\frac{3}{2}}\left(s, M_{K}^{2}, M_{\pi}^{2}\right)\nonumber\\[0.1cm]
\times&|F_{T}(s)|\,|F_{+}(s)|\,\sin\left[\delta_T(s)-\delta_+(s)\right]\,,
\end{align}
and 
\begin{align}\label{eq:finalACP}
& A_i^{CP}\simeq\Delta_{K\pi}\,S_\mathrm{EW}\,\frac{N_s}{n_i}\int_{s_{1,i}}^{s_{2,i}}\Bigg\{-\frac{\mathrm{Im}[\hat{\epsilon}_S]\mathrm{Im} \left[F_+(s)F_0^\ast(s)\right]}{m_\tau(m_s-m_u)}- \frac{2\mathrm{Im}[\hat{\epsilon}_T]\mathrm{Im}\left[F_T(s)F_0^\ast(s)\right]}{m_\tau}\,\nonumber\\[0.1cm]
&+\left[\left(\frac{1}{s}\!+\!\frac{\mathrm{Re}[\hat{\epsilon}_S]}{m_\tau(m_s\!-\!m_u)}\right)\mathrm{Re}\left[F_+(s)F_0^\ast(s)\right]\!-\!\frac{2\mathrm{Re}[\hat{\epsilon}_T]\mathrm{Re}[F_T(s)F_0^\ast(s)]}{m_\tau}\right]A^{CP}_K\Bigg\}C(s)ds\,,
\end{align}
where $s_{K\pi}=(M_K+M_{\pi})^2$ denotes the threshold of the $K\pi$ invariant mass squared, and $F_+(s)$, $F_0$, and $F_T(s)$ denote the $K\pi$ vector, scalar, and tensor form factors, respectively. $\delta_{+}(s)$ and $\delta_{T}(s)$ stand for the phases of the $K\pi$ vector and tensor form factors, respectively. The function $C(s)$ accounts for the detector efficiencies as well as all the model-independent terms. Here we adopt a seventh-order polynomial parametrization form of it as given in the supplementary material of Ref.~\cite{Bischofberger:2011pw}. 

\section{Numerical results and discussions}
\label{sec:numerical}

\subsection{Constraints on the NP parameters}
\label{sec:modelindepend}
 
\begin{figure}[ht]
	\centering
	\includegraphics[width=0.42\textwidth]{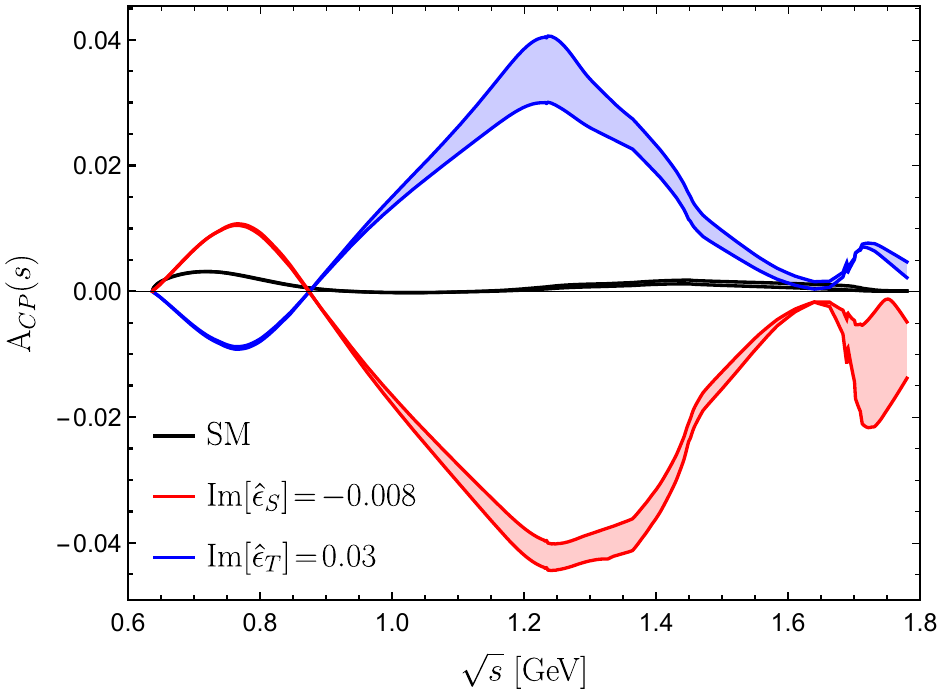}
	\hspace{0.12in}
	\includegraphics[width=0.42\textwidth]{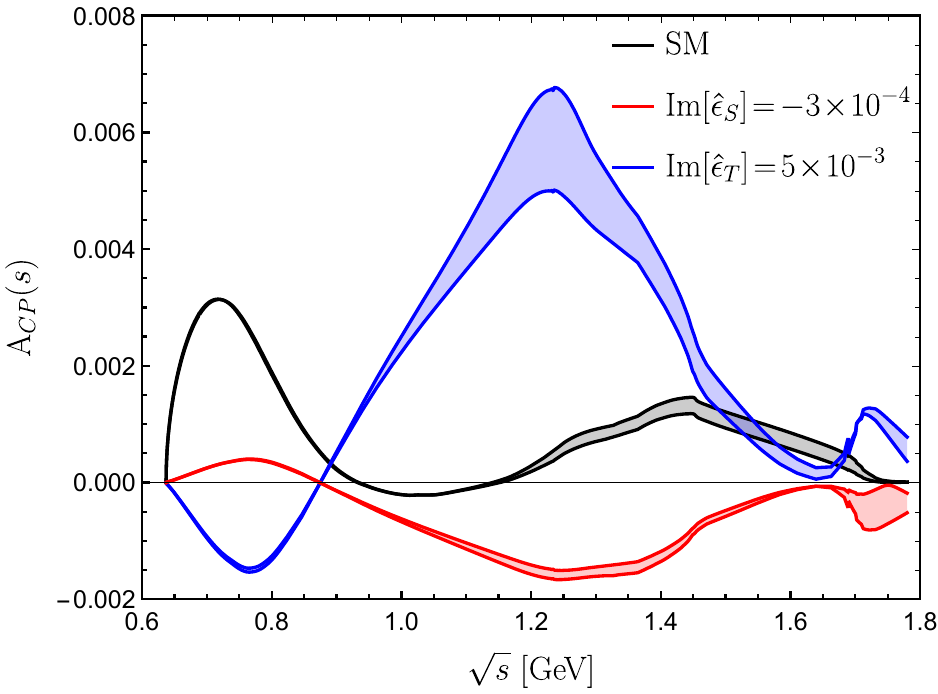}
	\caption{\small Left: Distributions of the $CP$ asymmetries in the whole $K\pi$ invariant-mass range for three different cases, the SM prediction (gray band), the non-standard scalar contribution with $\mathrm{Im}[\hat{\epsilon}_S]=-0.008$ (red band), and the non-standard tensor contribution with $\mathrm{Im}[\hat{\epsilon}_T]=0.03$ (blue band). Right: The zoomed-in version of the SM prediction as well as the cases with $\mathrm{Im}[\hat{\epsilon}_S]=-3\times10^{-4}$ (red band) and $\mathrm{Im}[\hat{\epsilon}_T]=5\times10^{-3}$ (blue band). \label{fig:ACP} }
\end{figure}

Combining the constraints from the branching ratio $\mathcal{B}_\mathrm{exp}^{\tau^-}=(4.04\pm0.02\pm0.13)\times10^{-3}$ and the CP asymmetries $A_{\mathrm{exp},i}^{CP}$ measured at four different bins by Belle, we can obtain the best fit values for the parameters $\mathrm{Im}[\hat{\epsilon}_S]$ and $\mathrm{Im}[\hat{\epsilon}_T]$ as
\begin{align}\label{eq:results}
\mathrm{Im}[\hat{\epsilon}_S]=-0.008\pm0.027, \qquad \qquad \mathrm{Im}[\hat{\epsilon}_T]=0.03\pm0.12\,.
\end{align}
In order to compare the NP contributions with the SM expectation for the CP asymmetries in the angular distributions of $\tau\to K_S\pi\nu_\tau$ decays, we also plot in Figure~\ref{fig:ACP} the distributions of the $CP$ asymmetries in the whole $K\pi$ invariant-mass range, with three different cases: the SM prediction induced by the indirect CP asymmetry in $K^0-\bar{K}^0$ mixing (gray band)~\cite{Chen:2020uxi}, the non-standard scalar contribution with the best-fit value $\mathrm{Im}[\hat{\epsilon}_S]=-0.008$ (red band), and the non-standard tensor contribution with the best-fit value $\mathrm{Im}[\hat{\epsilon}_T]=0.03$ (blue band). It can be seen that the CP asymmetries in the angular distributions of the decays could be significantly enhanced if these kinds of NP contributions are present.

\subsection{Bounds on the NP parameters from other processes}
\label{subsec:otherbounds}

\begin{figure}[ht]
	\centering
	\includegraphics[width=0.4\textwidth]{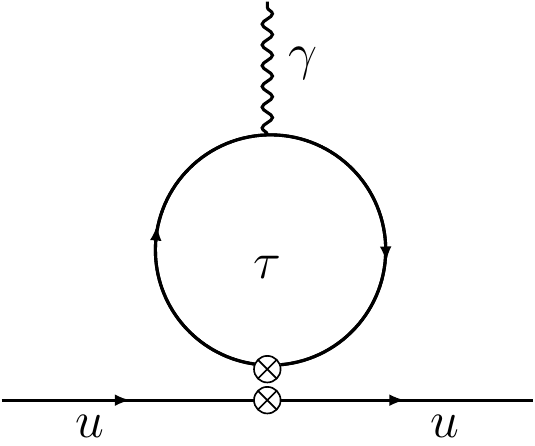}
	\hspace{0.3in}
	\includegraphics[width=0.40\textwidth]{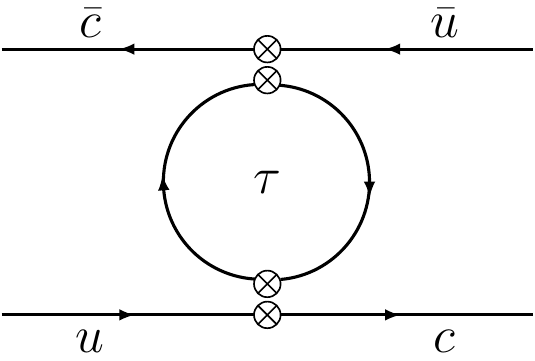}
	\caption{\small Diagrammatic representations of the electromagnetic dipole operator contributing to the neutron EDM induced by inserting the operator $(\bar{\tau}_L\sigma_{\mu\nu}\tau_R)(\bar{u}_L\sigma^{\mu\nu}u_R)$ (left), and the $\Delta C=2$ four-quark operators contributing to the $D^0-\bar{D}^0$ mixing produced by a double insertion of the operator $(\bar{\tau}_L\sigma_{\mu\nu}\tau_R)(\bar{c}_L\sigma^{\mu\nu}u_R)$ (right). \label{fig:nEDM&Dmixing} }
\end{figure}

\begin{figure}[ht]
	\centering
	\includegraphics[width=0.85\textwidth]{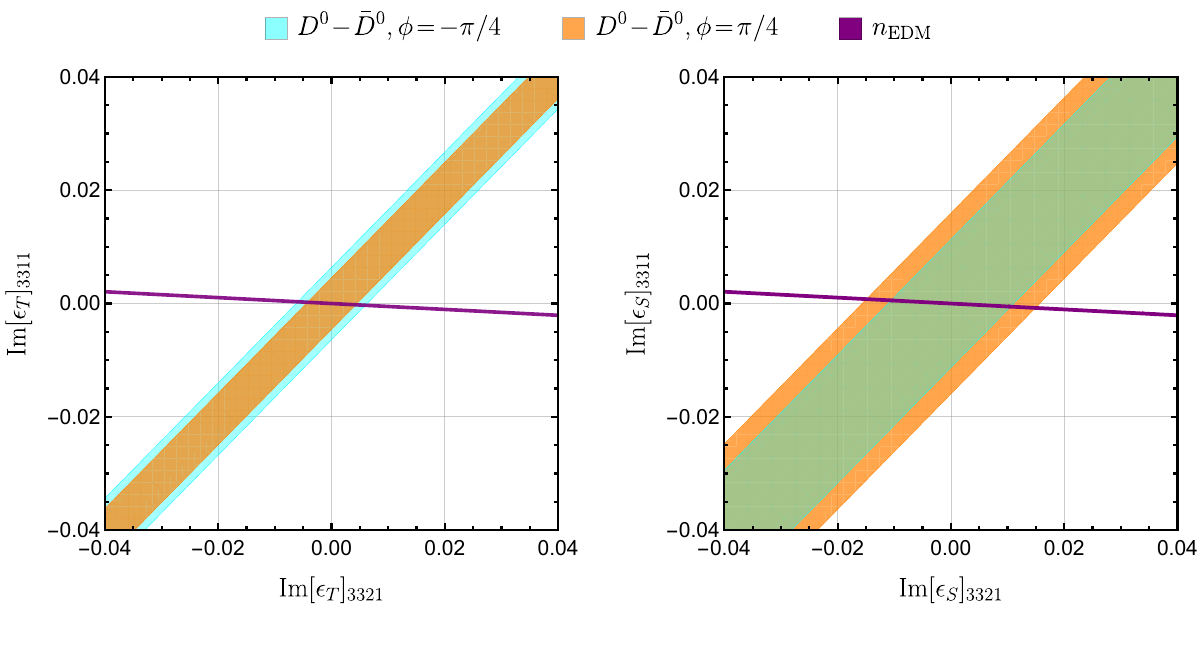}
	\vspace{-0.4cm}
	\caption{\small Allowed regions for the imaginary parts of the tensor (left) and scalar (right) coefficients from the neutron EDM (purple) and the $D^0-\bar{D}^0$ mixing with $\phi=-\pi/4$ (cyan) and $\phi=\pi/4$ (orange), where the NP scale is taken at $\Lambda=1~\text{TeV}$. Here $\phi$ is the phase of $V_{ud}V_{cd}\,[\epsilon_{T(S)}]_{3311}+V_{us}V_{cs}\,[\epsilon_{T(S)}]_{3321}$, and the choices $\phi=\pm\pi/4$ are to ensure that the NP Wilson coefficients $C^\prime_{2,3}$ are purely imaginary~\cite{Cirigliano:2017tqn}. \label{fig:com} }
\end{figure}

If one assumes that the NP stems from physics well above the electroweak scale, the LEFT Lagrangian in Eq.~\eqref{eq:LEFT} shall then be matched onto the following $SU(2)_L$ invariant SMEFT Lagrangian~\cite{Buchmuller:1985jz,Grzadkowski:2010es,Brivio:2017vri},
\begin{align}\label{eq:SMEFTLT}
&\mathcal{L}_\mathrm{SMEFT} \supset [C^{(3)}_{\ell equ}]_{klmn} (\bar{\ell}^i_{Lk}\sigma_{\mu\nu}e_{Rl})\epsilon^{ij}(\bar{q}_{Lm}^{j}\sigma^{\mu\nu}u_{Rn})\, \nonumber \\[0.2cm]
&+[C^{(1)}_{\ell equ}]_{klmn}(\bar{\ell}^i_{Lk}e_{Rl})\epsilon^{ij}(\bar{q}_{Lm}^{j}u_{Rn})+[C_{\ell edq}]_{klmn}(\bar{\ell}^i_{Lk}e_{Rl})(\bar{d}_{Rm}q_{Ln}^{i})+{\rm h.c.}\,.
\end{align}
Rewriting the Lagrangian in the mass basis, we can see that the same SMEFT operators contributing to $\tau\to K_S\pi\nu_\tau$ decays also contribute to the neutron EDM and the $D^0-\bar{D}^0$ mixing~\cite{Cirigliano:2017tqn}, as shown in Figure~\ref{fig:nEDM&Dmixing}. The latter provide very stringent limits on the imaginary parts of the non-standard scalar and tensor coefficients $\mathrm{Im}[\epsilon_{S,T}]_{3321(3311)}$, as shown in Figure~\ref{fig:com}. Under the ``single coefficient dominance'' assumption, we find that (i) $|\mathrm{Im}[\epsilon_{T}]|<4\times10^{-6}$, which is five orders of magnitude smaller than the value required to explain the $2.8\sigma$ discrepancy between the BaBar measurement and the SM prediction of $A_\mathrm{rate}^{CP}$; (ii) $|\mathrm{Im}[\epsilon_{S}]|<2.3\times10^{-3}$, which is comparable with that obtained from the $\tau\to K_S\pi\nu_\tau$ decays. However, when there exist extraordinary cancellations in the combinations $V_{ud}^2\mathrm{Im}[\epsilon_{T(S)}]_{3311}+V_{us}^2\mathrm{Im}[\epsilon_{T(S)}]_{3321}$ (for the neutron EDM) as well as $V_{ud}V_{cd}\mathrm{Im}[\epsilon_{T(S)}]_{3311}+V_{us}V_{cs}\mathrm{Im}[\epsilon_{T(S)}]_{3321}$ (for the $D^0-\bar{D}^0$ mixing), the bounds on $|\mathrm{Im}[\epsilon_{T(S)}]|$ could be significantly diluted~\cite{Chen:2021udz}. 

We can therefore conclude that, once the bounds from the neutron EDM and the $D^0-\bar{D}^0$ mixing are taken into account, neither the scalar nor the tensor interaction can produce any significant effects on the CP
asymmetries in the decays considered, especially under the ``single coefficient dominance'' assumption. Nevertheless, when there exist extraordinary cancellations between the NP contributions, the non-standard scalar and tensor interactions can still produce observable effects on the CP asymmetries in the angular distributions of $\tau\to K_S\pi\nu_\tau$ decays.

\section{Conclusion}

Here we have presented a detailed study of the CP asymmetries in the decay rate and angular distributions of $\tau\to K_S\pi\nu_\tau$ decays. In the SM, both of them are non-zero due to the CPV effect in $K^0-\bar{K}^0$ mixing. The BSM CPV contributions were then investigated in a model-independent way by using the LEFT framework. The resulting bounds on the imaginary parts of the non-standard scalar and tensor couplings are given, respectively, by $\mathrm{Im}[\hat{\epsilon}_S]=-0.008\pm0.027$ and $\mathrm{Im}[\hat{\epsilon}_T]=0.03\pm0.12$. Using the obtained best-fit values, we have also presented the distributions of the CP asymmetries, finding that significant deviations from the SM prediction are possible in almost the whole $K\pi$ invariant-mass range. Therefore, the CPV angular observables considered here are an ideal probe of the non-standard scalar and tensor interactions. While being still plagued by large experimental uncertainties, the current constraints obtained will be improved with more precise measurements from the Belle II experiment~\cite{Kou:2018nap}, as well as the future Tera-Z~\cite{Pich:2020qna} and STCF~\cite{Sang:2020ksa,Achasov:2023gey} facilities.

If the NP originates from above the electroweak scale, the $SU(2)_L$ invariance of the SMEFT Lagrangian would indicate that very strong limits on the imaginary parts of the non-standard scalar and tensor coefficients can also be obtained from the neutron EDM and $D^0-\bar{D}^0$ mixing. We found that, unless there exist extraordinary cancellations between the NP contributions, neither the scalar nor the tensor interaction can produce any significant effects on the CP asymmetries in the decays considered, especially under the ``single coefficient dominance'' assumption.

\section*{Acknowledgements}
This work is supported by the NNSFC under Grant Nos. 12135006 and 12075097, as well as the Fundamental Research Funds for the Central Universities under Grant Nos. CCNU22LJ004 and 11623330. FC is also supported by the 2024 Guangzhou Basic and Applied Basic Research Scheme Project for Maiden Voyage (2024A04J4190).

\bibliographystyle{ws-ijmpa}
\bibliography{reference}
\end{document}